\documentclass[a4paper,onecolumn,11pt,unpublished
]{quantumarticle}
\pdfoutput=1
\usepackage{epsfig,amssymb,amsmath,mathrsfs,amsthm,graphicx,float,xcolor,hyperref}

\usepackage[numbers,sort&compress]{natbib}
\usepackage{comment}
\usepackage{subfigure}
\usepackage{algorithm}
\usepackage{algpseudocode}
\usepackage{refcount}

\newcommand{\N}{\mathbb{N}}
\newcommand{\Z}{\mathbb{Z}}

\newcommand{\spc}[1]{\mathcal{#1}}

\def\d{{\rm d}}

\def\>{\rangle}
\def\<{\langle}

\newcommand{\map}[1]{\mathcal{#1}}
\newcommand{\Tr}{\operatorname{Tr}}

\floatname{algorithm}{Protocol}

\newtheorem{lemma}{Lemma}

\newtheorem{cor}{Corollary}
\newtheorem{defi}{Definition}

\def\qed{$\blacksquare$ \newline}

\newcommand*{\YY}[1]{{\color{red} [YY: #1]}}

\usepackage{tikz}
\usepackage{lipsum}

\newtheorem{theorem}{Theorem}

\begin{document}

\title{Ultimate limit on time signal generation} 

\author{Yuxiang Yang} \affiliation{Institute for Theoretical Physics, ETH Z\"urich, Switzerland}\orcid{0000-0002-0531-8929}\email{yangyu@ethz.ch} 

\author{Renato Renner} \affiliation{Institute for Theoretical Physics, ETH Z\"urich, Switzerland}\orcid{0000-0001-5044-6113}\email{renner@ethz.ch}

\maketitle

\begin{abstract} 
The generation of time signals is a fundamental task in science.  Here we study the relation between the quality of a time signal and the physics of the system that generates it. According to quantum theory, any time signal can be decomposed into individual quanta that lead to single detection events. Our main result is a bound on how sharply peaked in time these events can be, which depends on the dimension of the signal generator. This result promises applications in various directions, including information theory, quantum clocks, and process simulation.
\end{abstract}

\maketitle

\section{Introduction: What is required to generate a desired time signal?}

\begin{figure}  [h!]
\begin{center}
  \includegraphics[width=0.95\linewidth]{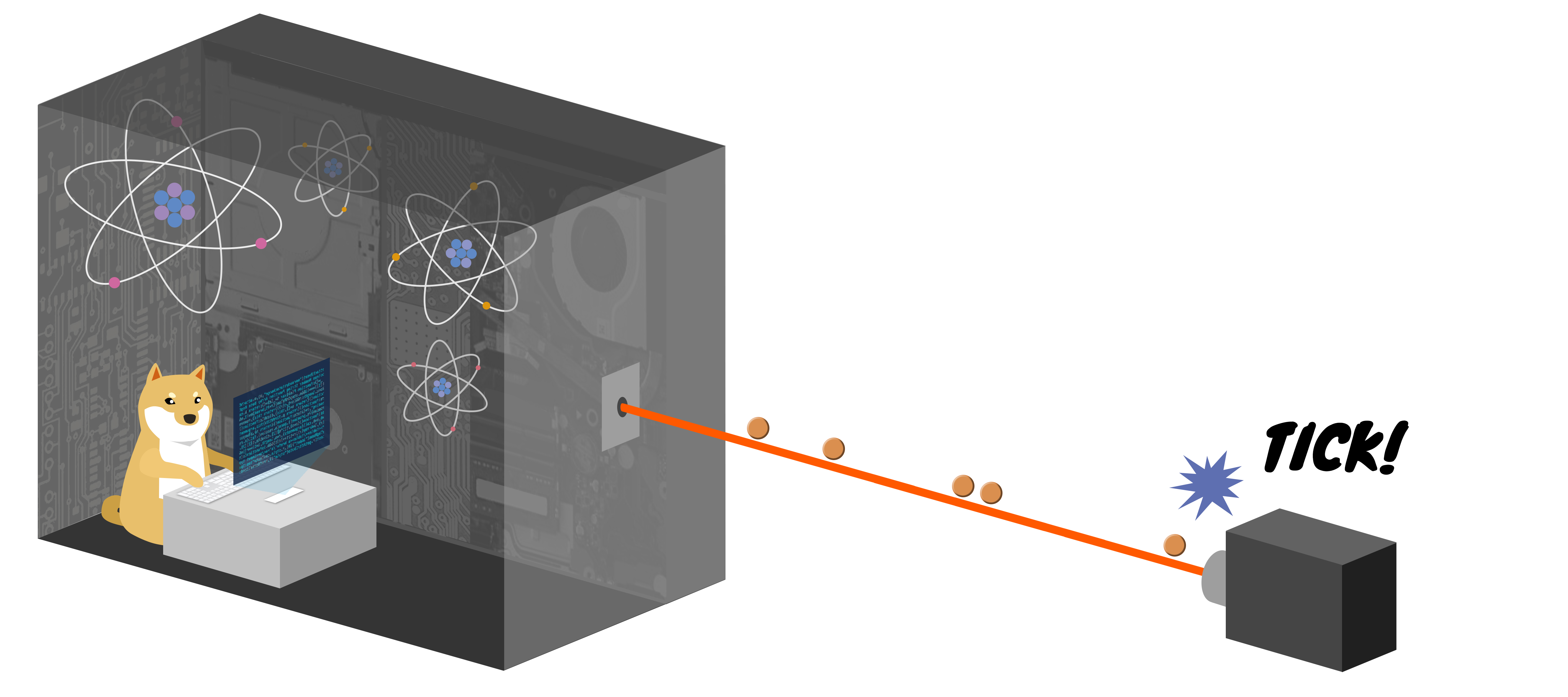}
  \end{center}
\caption{\label{fig-setup}
  {\bf Quantised view on signal generation.} A signal generator with potentially sophisticated inner structure (on the left) emits a beam of quanta as a time signal. They hit a detector (on the right) sequentially. We are interested in the time distribution of the detection events (the ``ticks'').}
\end{figure}

Physical processes that generate time signals are ubiquitous.
Signal sources generate the pulses that drive our electronic devices, neurons fire trains of spikes, and clocks tick. Understanding the relation between a signal and its source is a core task in many scientific disciplines.
From the observation of a signal one seeks insight into the, potentially complex, mechanism that generates it (see Figure~\ref{fig-setup}). This path of research has been widely taken in modern physics, astronomy, electrical engineering, biology, and computer science. 
A basic question is thus: given a time signal, characterised by a function over time, what are the necessary physical requirements on the source, i.e., the dynamical system that generates it?

Conventionally, a time signal consists of a large population of quanta, possibly generated by oscillators and then further modulated by filtering out some of them. 
A classical example would be an apparatus that consists of an oscillator followed by a diode clipping circuit, which cuts off part of the harmonic signal by limiting the number of quanta that can pass through it per unit of time.
Fundamentally, a time signal generator is specified by a model that describes how each of the quanta is generated. 
We may thus focus on single-quantum generators, i.e., machines that generate a signal that leads to one single detection event. These machines can then be regarded as the fundamental building blocks for any general signal generator.
The most powerful single-quantum generator would be one which produces a quantum at a well-defined moment in time, i.e., whose detection probability distribution over time is a delta-function.

However, an idealised single-quantum generator is not feasible: the generation of an event with a delta-distribution over time requires an infinite-dimensional machine and is thus unphysical \cite{pauli1933handbuch}.  
Notice that dimension here refers to the degrees of freedom that can be well controlled, which is a scarce resource in engineering quantum devices. (For realistic systems,  this \emph{effective} dimension is usually small and always finite, as discussed below.) 
One would then turn attention to generating good approximations of delta-signals, i.e., signals that are \emph{sharp enough}.
We stress that these play a fundamental role: If we know how to generate single-quantum signals that approximate  delta-functions, we can approximate arbitrary time signals with good precision. 

To produce an approximate delta-function at any designated time, the generator needs the capability of remaining silent for a long time and then producing a peaked signal. The sharpness of the signal is thus captured by both the duration of the silence before the peak and the width of the peak. 
For a single-particle time signal, we can measure its \emph{sharpness} by \cite{woods2018quantum}
\begin{align}
R:=\frac{\mu^2}{\sigma^2},
\end{align}
where $\mu$ and $\sigma$ are the mean and the standard deviation of the generated distribution, respectively. The higher $R$ is, the more similar the distribution is to a delta-function. 

To analyse the cost of generating a sharp time signal, we need to quarantine the generator from any other source of time information. 
This, in particular, excludes any time-dependent dynamics, for it would mean that the system is actually driven by another signal and is thus forbidden. 
Technically, the dynamics of the generator has to be \emph{Markovian and time-homogeneous}. 
To compensate for this constraint, but sill be able to exhibit the desired time-inhomogeneous behaviour, i.e., staying silent for a long time and then firing all of a sudden, the device needs its own memory. Indeed, if the generator has no memory, meaning that its dimension is one, it can only generate  Poisson point processes, which are completely random and are very non-sharp.
Therefore, intuitively the dimension of a generator and the sharpness of the distribution it generates are dual to each other.

In this work, we make this intuition quantitatively precise by establishing the following bound (see Theorem~\ref{theorem-main} for more details): For large $d$, the sharpness $R$ of any $d$-dimensional signal generator scales as 
\begin{align}\label{rough-bound}
R\le 2\pi e\cdot d^2.
\end{align}
The bound does not only hold for perfectly controlled quantum generators, but also for arbitrary signal generators, including macroscopic classical generators. For the latter, the dimension $d$ should be replaced by its \emph{controllable dimension} $d_{\rm ctrl}$, which is, intuitively, the number of mutually distinguishable states that the generator can reach while evolving. A pendulum, for instance, is a macroscopic object whose ``quantum dimension'' $d$ is either infinite or undefinable. However, because its motion is restricted to a collective degree of freedom (namely the centre of mass motion), its controllable dimension $d_{\rm ctrl}$ is a finite number that depends on its mass (number of atoms). We can thus compare it to a microscopic quantum generator. In particular, we show that, fixing the number of atoms, gaining individual control over each of the degrees of freedom offers an exponential boost of performance.
This fact may for the moment only be of theoretical interest, but will become relevant to building advanced clocks once we can control $30-40$ (logical) qubits with high fidelity (possibly with the use of techniques from fault-tolerant computing).

The remaining part of this paper is organised as follows: In Section~\ref{sec-model} we introduce a quantum model of signal generators, and in Section~\ref{sec-bound} we state and prove our main result.
In Section~\ref{sec-intuition} we extend the bound to generators that are not perfectly controlled, by generalising the concept of dimension to that of a controllable dimension $d_{\rm ctrl}$. 
In Section~\ref{sec-discussion} we discuss the connection between our result and  
various current research directions. Finally, we conclude with a brief discussion on future perspectives in Section~\ref{sec-conclusion}.

\section{Modelling time signal generators}\label{sec-model}
In this section,  we introduce the general model to describe time signal generators. Note that equivalent models have been used to define autonomous quantum clocks \cite{rankovic2015quantum,woods2018quantum}. 

As mentioned in the introduction, the core of a signal generator is a quantum system with time-homogeneous Markovian dynamics. A simple example of such dynamics is the unitary evolution generated by a time-independent Hamiltonian. However, such a signal generator is trivial, 
since its evolution would correspond to that of an isolated system. It would thus remain silent and never output anything unless an external party makes a measurement on it (which obviously violates requirement of homogeneity in time). Therefore, we need to consider generic open system dynamics.

\begin{figure}  [b!]
\begin{center}
  \includegraphics[width=0.9\linewidth]{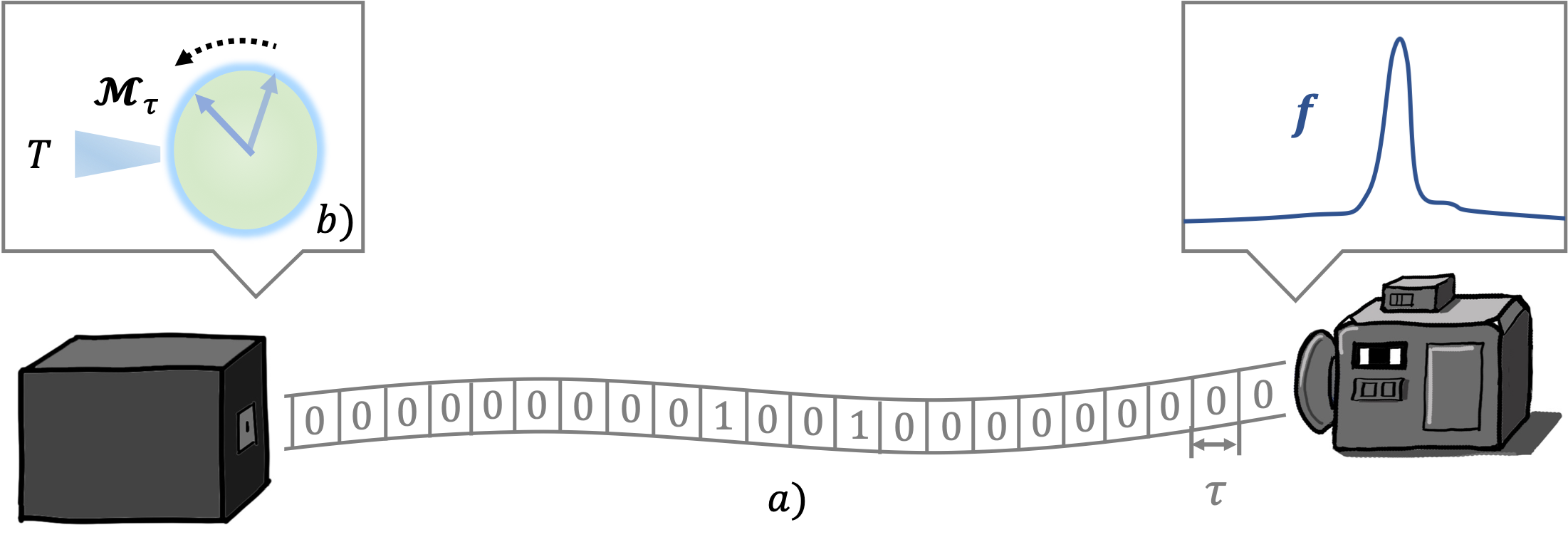}
  \end{center}
\caption{\label{fig-generator}
   {\bf $a)$ Time signal generator and its output.} A time signal generator outputs an infinitely long tape consisting of a sequence of tick registers. Each of them can be in a ``vacuum state ($|0\>$) or contain a ``tick" ($|1\>$). One considers the limit where the time $\tau$ during which the generator operates on a single tick register is arbitrarily small. The tape does not need to contain a classical string, but may be in coherent superposition of different strings. The time signal is characterised by a function $f(t)$, which is the density of ticks observed by a detector. {\bf $b)$ Inside the generator.} In a generator, a $d$-dimensional quantum system $S$ evolves under dynamics $\map{M}_\tau:S\to ST$ and outputs information to the current tick register~$T$.}
\end{figure}

A time-homogeneous Markovian dynamics on a quantum system $S$ is characterised by a dynamical semigroup $\{\overline{\map{M}}_{t}\}_{t}$, where $\overline{\map{M}}_{t}$ for every $t\ge0$ is a quantum channel (a completely positive trace-preserving linear map) acting on $S$ such that $\overline{\map{M}}_{t_1}\overline{\map{M}}_{t_2}=\overline{\map{M}}_{t_1+t_2}$ for any $t_{1,2}$ and $\lim_{t\to 0^+}\overline{\map{M}}_{t}=\map{I}_S$. Since $\overline{\map{M}}_{t}$ is a generic quantum channel instead of a unitary, it allows the generator's system to \emph{emit information to the outside world}, and thus the generation of a nontrivial signal is possible.

To describe the emission of time signals we consider a sequence of two-dimensional registers $T_1, T_2, T_3,\dots$, called ``tick registers'', that sequentially interact with the clock. 
Each tick register has a state $|0\>$ standing for ``vacuum'' and a state $|1\>$ standing for a tick (physically a photon, for instance).
The above description of the dynamics may then be extended to a family of quantum channels $\{\map{M}_{\tau}:S\to ST\}_{\tau\ge 0}$ that characterises the infinitesimal (i.e.\ arbitrarily small $\tau$) behaviour of the signal generator, where $T$ stands for any of the tick registers.
Within a very short time $\tau$, the generator evolves and consults its system $S$ whether to emit a tick, which is then written into the current tick register. The time signal corresponds to the resulting population of ticks in the tick registers
(see Figure~\ref{fig-generator}).

The dynamics on $S$ is thus $\Tr_T\,\map{M}_\tau$ for infinitesimal $\tau$. The relation between $\map{M}$ and $\overline{\map{M}}$, which is the aforementioned dynamical semigroup, can be established as
\begin{align}\label{dynamics_S}
\overline{\map{M}}_t:=\lim_{\tau\to 0^+}\left(\Tr_T\,\map{M}_\tau\right)^{\lfloor\frac{t}{\tau}\rfloor},
\end{align}
where $\lfloor\cdot\rfloor$ is the floor function.
That is, $\Tr_T\,\map{M}_\tau$ defines the instantaneous evolution, whereas $\overline{\map{M}}_t$ reflects the system's dynamics for arbitrary time intervals.
From this definition [Eq.\ (\ref{dynamics_S})], to ensure that $\{\overline{\map{M}}_t\}$ is Markovian and time-homogeneous, we only need to assume
\begin{align}\label{generator-assumption}
\lim_{t\to 0^+}\lim_{\tau\to 0^+}\left(\Tr_T\,\map{M}_\tau\right)^{\lfloor\frac{t}{\tau}\rfloor}=\map{I}_S.
\end{align}
Therefore, a generic time signal generator can be defined as follows:
\begin{defi}[Time signal generators]\label{defi-generator}
A time signal generator is characterised by a tuple $\left(\rho_0,\left\{\map{M}_{t}\right\}\right)$, where $\rho_0$ is the initial state of a $d$-dimensional (quantum) system $S$ and $\{\map{M}_\tau:S\to ST\}_{\tau\ge 0}$ is a family of quantum channels satisfying Eq.\ (\ref{generator-assumption}). $T$ is called a tick register and assumed to be equipped with an orthonormal basis $\{|0\>,|1\>\}$.
\end{defi}

In principle, even for arbitrarily small $\tau$, the channel $\map{M}_{\tau}$ is allowed to generate any state on the tick register $T$.  We will usually assume that the machine operates on tick registers that are initialised to their vacuum states. The quantum speed limit \cite{mandelstam1991uncertainty} limits the change of state in $\tau\ll 1$. Applied to our description, this means
\begin{align}\label{qsl}
\Tr_S\left(\lim_{\tau\to0^+}\map{M}^{S\to ST}_{\tau}(\rho)\right)=|0\>\<0|_T
\end{align}
for any $\rho$ on the system $S$.

Now we can look at the signal generated by the generator. As shown in Figure~\ref{fig-generator}, starting with $S$ in state $\rho_0$, the signal is generated by consequentially applying $\map{M}_{\tau}^{S\to ST_1}, \map{M}_{\tau}^{S\to ST_2}, \dots$ (here the superscript $S\to ST_i$ is introduced to distinguish between the different tick registers). The signal thus consists of infinite strings of the type ``$\dots |01001000\>\dots$" with each $|1\>$ standing for a tick. Usually, under the conditions leading to Eq.\ (\ref{qsl}), in the limit of small $\tau$ there will be many more zeros than ones in the  string.

The output string of a signal generator can in principle exhibit quantum coherence. However, the signal is observed when the tick registers hit a detector. The detector responds to every incoming tick by amplifying it to a classical signal that can be observed.
The cumulative function of the signal, which quantifies how many ticks are expected in $[0,t)$, can be expressed as:
\begin{align}
F\left(t\right):=\lim_{\tau\to0}\sum_{\vec{n}\in\{0,1\}^{\lfloor\frac{t}{\tau}\rfloor}}|\vec{n}|\cdot\Tr\left[\left(I_S\otimes|\vec{n}\>\<\vec{n}|\right)\left(\bigcirc_{i=1}^{\lfloor\frac{t}{\tau}\rfloor}\map{M}_{\tau}^{S\to ST_i}(\rho_0)\right)\right],
\end{align}
where $|\vec{n}|$ is the number of ones in the bit string $\vec{n}\in\{0,1\}^{\lfloor\frac{t}{\tau}\rfloor}$.
The density function of the signal is then defined via the derivative of the cumulative function, i.e.
\begin{align}
f(t):=\left.\frac{\d F(x)}{\d x}\right|_{x=t}.
\end{align}
We remark that, for the functions $F(t)$ and $f(t)$, coherence between different output strings does not play a role.

An important category of generators are the \emph{singleton generators} that produce one and only one tick. 
Technically, this requires 
\begin{align}\label{single-generator-req}
\lim_{t\to\infty}\lim_{\tau\to0}\sum_{\vec{n}\in\{0,1\}^{\lfloor\frac{t}{\tau}\rfloor},|\vec{n}|\not=1}\Tr\left[\left(I_S\otimes|\vec{n}\>\<\vec{n}|\right)\left(\bigcirc_{i=1}^{\lfloor\frac{t}{\tau}\rfloor}\map{M}_{\tau}^{S\to ST_i}(\rho_0)\right)\right]=0
\end{align}
as well as the cumulative function to satisfy
\begin{align}\label{acc-unit}
\lim_{t\to\infty}F(t)=1.
\end{align}
Then the density function $f(t)$ is a probability density function (pdf), and we can properly define the sharpness of its signal.  
\begin{defi}[Singleton generators]\label{defi-single-generator}
A singleton generator is a time signal generator (cf.\ Definition~\ref{defi-generator}) such that Eqs.\ (\ref{single-generator-req}) and (\ref{acc-unit}) hold.
The sharpness of a singleton generator's signal, which has a pdf $f(t)$, is defined as
\begin{align}\label{R}
R:=\frac{\mu^2}{\sigma^2},
\end{align}
where  $\mu:=\int_0^{\infty}\d t\,f(t)t$ and  $\sigma^2:=\int_0^{\infty}\d t\,f(t)(t-\mu)^2$ are the mean and the variance, respectively.
\end{defi}

Any generator $(\rho_0,\{\map{M}_\tau\})$ that produces at least one tick can be turned into a singleton generator by forcing it to remain silent after producing its first tick: We first introduce a silent state $\psi_{\rm silent}$ and enlarge the system's Hilbert space to $\spc{H}_S\oplus\psi_{\rm silent}$ (this increases the dimension by only one). Then we modify the dynamics to $\map{M}_{\rm switch}(\map{M}_{\tau}\oplus\map{M}_{\rm silent})$. Here $\map{M}_{\rm switch}$ acts nontrivially only if the state of the tick register is $|1\>\<1|$. In this case, $\map{M}_{\rm switch}$ sets the state of the system to $\psi_{\rm silent}$. Meanwhile, $\map{M}_{\rm silent}:=\map{I}_{\rm silent}\otimes|0\>\<0|_T$ acts trivially on the system and never produces a tick.
Conversely, we can build a multi-signal generator by concatenating singleton generators.

\section{Lower bound on the dimension}\label{sec-bound}
\subsection{Statement of the main result}
In the following we state and then prove our main result [Eq.\ (\ref{rough-bound})].
Recall that our task is to see how well a $d$-dimensional signal generator can produce a signal that approximates a delta-signal, which contains only one tick and whose sharpness is measured by $R$.
\begin{theorem}\label{theorem-main}
For any $c>2\pi e$, there exists a $d_0$ such that for any singleton generator (cf.\ Definition~\ref{defi-single-generator}) of dimension $d>d_0$, the sharpness of its signal obeys the bound 
\begin{align}\label{bound}
R\le c\cdot d^2.
\end{align} 
\end{theorem}
Explicitly, we will prove that for any singleton generator of dimension $d\ge 4$ the following inequality holds:\footnote{We shall use the convention $\log:=\log_2$ throughout the paper.}
\begin{align}\label{bound-explicit}
\log d\ge& \left(1-30d^{-\frac12}\right)\left(\frac12\log\frac{R}{2\pi e}\right)-\epsilon_d\\
&\qquad\epsilon_d:=d^{-\frac12}\log\left(d^8e^{\frac52}(2\pi e)^{16}\right)-12(\log e)d^{-\frac52}-2\zeta_{d^{\frac{13}{4}}/2}.
\end{align}
Here $\zeta_{d^{\frac{13}{4}}/2}$ [see later Eq.\ (\ref{zeta-sigma}) for the concrete expression] is an error term   
that decreases exponentially in $\pi^2 d^{\frac{13}{2}}/2$ in the asymptotic limit of large $d$. Therefore, there exists $d_0$ (potentially dependent on $c$) such that Eq.\ (\ref{bound}) holds for any $d> d_0$.


\subsection{Definitions for the proof}

Before the main course of the proof, we prepare a couple of necessary definitions.
First, we introduce a coarse-graining parameter $\Delta$, which should always be a vanishing function of $d$. Explicitly, here we choose 
\begin{align}\label{Delta}
\Delta:=\left(\frac{1}{R^2}\right)\cdot\left(\frac{\mu}{d}\right).
\end{align}
The above choice implies that
\begin{align}\label{Delta-mu-sigma}
\frac{\sigma}{\Delta}=d\cdot R^{\frac32}\qquad{\rm and}\qquad\frac{\mu}{\Delta}=d\cdot R^2.
\end{align}
We now introduce an important definition concerning the dynamics of the generator within a period $\Delta$, with the tick registers coarse-grained. Explicitly, we define a quantum channel $\map{M}_\Delta: S\to ST$, with $T$ being a (classical) bit register, as
\begin{align}
\map{M}_\Delta(\rho):=(\map{I}_S\otimes\map{C})\lim_{\tau\to0^+}\left(\bigcirc_{i=1}^{\lfloor\frac{t}{\tau}\rfloor}\map{M}_{\tau}^{S\to ST_i}\right),
\end{align}
where $\map{C}$ acts on all the tick registers $T_1,T_2,\dots$ to coarse-grain the qubit string to one single bit, depending on whether a tick is contained in the string. Then the new tick register $T$ is a single-bit register that shows whether any tick is produced during $[0,\Delta)$ or not.
We take ``snapshots'' of the generator system conditioned on no tick production, by defining the following quantum state:
\begin{align}\label{rho_k}
\rho^\Delta_k:=\frac{\Tr_{T_1\cdots T_k}\left[(I_S\otimes|0\>\<0|_{T_1}\otimes\cdots\otimes|0\>\<0|_{T_k})\left(\map{M}_\Delta\right)^k(\rho_0)\right]}{P^\Delta(k)},
\end{align}
where $\rho_0$ is the initial state, $(\map{V})^k$ is the map $\map{V}$ iterated for $k$ times, and 
\begin{align}
P^\Delta(k):=\Tr\left[(I_S\otimes|0\>\<0|_{T_1}\otimes\cdots\otimes|0\>\<0|_{T_k})\left(\map{M}_\Delta\right)^k(\rho_0)\right] 
\end{align}
is the probability that no tick has been generated until step $k$. By definition, $\{\rho^\Delta_k\}_{k=0}^{\infty}$ forms the trajectory of the machine state conditioned on no singleton production, with an interval of $\Delta$.

\begin{figure}  [h!]
\begin{center}
  \includegraphics[width=0.7\linewidth]{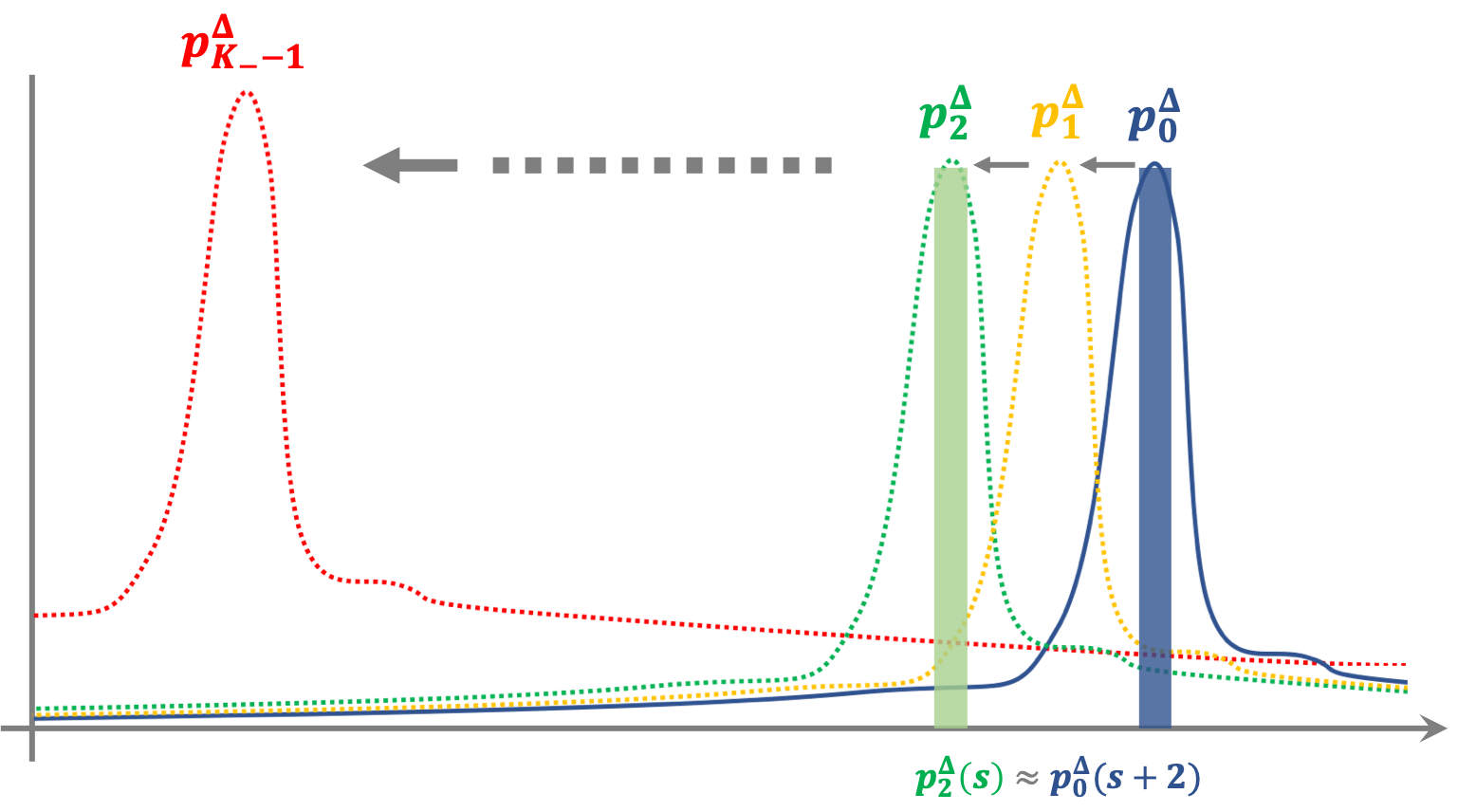}
  \end{center}
\caption{\label{fig-pdfs}
  {\bf Relation between $p_0^\Delta$ and $p_k^\Delta$ for any $k< K_-$.} By Eq.\ (\ref{relation_pdfs}), $p_{k}^\Delta$ is $p_0^\Delta$ left shifted for $k$ units. Therefore, if $P^\Delta(k)$ is close to one (which is the case if $k<K_-$), the $s$-th entry of $p_{k}^\Delta$ is close to the $(s+k)$-th entry of $p_0^\Delta$.}
\end{figure}

Next, we define the probability distribution of the tick, when the initial state of the machine is $\rho_k^\Delta$. This is given by the probability that a tick is produced at the $s$-th step of the evolution:
\begin{align}
p_{k}^\Delta(s):=\Tr\left[\left(I_{\rm S}\otimes|0\>\<0|_{T_1}\otimes\cdots\otimes|0\>\<0|_{T_{s-1}}\otimes|1\>\<1|_{T_s}\right)\left(\map{M}_\Delta\right)^{s}\left(\rho^\Delta_k\right)\right].
\end{align}
From the above definition, it is straightforward to see that $p_{0}^\Delta=p_{k=0}^\Delta$ is essentially the original pdf $f$, discretized with a step width of $\Delta$ (see Figure~\ref{fig-pdfs}).

Substituting Eq.\ (\ref{rho_k}) into the above definition, we get a useful relationship: For any $s$ and for any $k<s$,
\begin{align}
p^\Delta_0(s)&=P^\Delta(k)\Tr\left[\left(I_{\rm S}\otimes|0\>\<0|_{T_{k+1}}\otimes\cdots\otimes|0\>\<0|_{T_{s-1}}\otimes|1\>\<1|_{T_s}\right)\left(\map{M}_\Delta\right)^{s-k}\left(\rho^\Delta_k\right)\right]\nonumber\\ 
&=P^\Delta(k)\cdot p_{k}^\Delta(s-k).\label{relation_pdfs}
\end{align}
The relation essentially says that any $p_{k}^\Delta$ can be obtained by left-shifting $p_0^\Delta$ for $k$ steps.

\subsection{Proof of Theorem~\ref{theorem-main}}
Now we are ready to derive the bound on the sharpness $R:=\mu^2/\sigma^2$ of any distribution $f$ generated by a $d$-dimensional machine. For convenience of the proof, we assume $d\ge 4$ without loss of generality (as we can always set $d_0\ge 4$ in Theorem~\ref{theorem-main}).

We introduce a control register $C$, which is a $K_-$-dimensional classical system $\{|k\>\<k|\}_{k=0}^{K_--1}$. For now $K_-$ is an arbitrary positive integer, and later we will set it to be a function of $d$, $\Delta$, $\mu$, and $\sigma$.
Consider the following classical-quantum state:
\begin{align}\label{sigma0}
\sigma_{0}^{CS}:=\sum_{k=0}^{K_--1}
\tilde{P}^\Delta(k)|k\>\<k|^C\otimes \left(\rho^\Delta_k\right)^S.
\end{align}
where 
\begin{align}\label{CK_and_P_tilde}
\tilde{P}^\Delta(k):=P^\Delta(k)\cdot C^\Delta_{K_-}\qquad C^\Delta_{K_-}:=\left(\sum_{k=0}^{K_--1}P^\Delta(k)\right)^{-1}
\end{align}
is a probability distribution on $\{|k\>\<k|\}_{k=0}^{K_--1}$.
By applying $\left(\map{M}_\Delta\right)^N$ with $N\to\infty$ and tracing out the machine system on $\sigma_0$, we get the following bipartite distribution:
\begin{align}
\sigma_{1}^{CT^\infty}&:=\left(\map{I}_{C}\otimes\left(\Tr_S\circ\left(\map{M}_\Delta\right)^{\infty}\right)\right)\left(\sigma_{0}^{CS}\right)\nonumber\\
&=\sum_{k=0}^{K_--1}\tilde{P}^\Delta(k)|k\>\<k|^C\otimes\Tr_{S}\left(\map{M}_\Delta\right)^{\infty}\left(\rho_k^\Delta\right)\nonumber\\
&=\sum_{k=0}^{K_--1}\tilde{P}^\Delta(k)|k\>\<k|^C\otimes \left(p_k^\Delta\right)^{T^\infty},\label{sigma1}
\end{align}
where $T^\infty$ denotes the collection of all tick registers.
Here, abusing the notations a bit, we denote by $p_k^\Delta$ the classical state $\sum_{s=0}^{\infty}p_k^\Delta(s)|0\>\<0|^{\otimes (s-1)}\otimes|1\>\<1|\otimes|0\>\<0|^{\otimes \infty}$ on $T^\infty$.

By data processing and the dimension bound $I(A:B)\le \log d_B$ for the mutual information of a classical-quantum state, we have the following chain of inequalities:
\begin{align}
\log d\ge  I(C:S)_{\sigma_0}\ge I(C:T^{\infty})_{\sigma_1}.
\end{align}
Denote by $\<g_k\>:=\sum_k g_k \tilde{P}^\Delta(k)$ the average of a function $g_k$ with respect to the distribution $\tilde{P}^\Delta(k)$.
 Since $I(C:T^{\infty})_{\sigma_{1}}=H\left(\left\<p^\Delta_k\right\>\right)-\left\<H\left(p^\Delta_k\right)\right\>$, we get 
\begin{align}\label{dimension_vs_entropies_bound}
\log d\ge H\left(\left\<p^\Delta_k\right\>\right)-\left\<H\left(p^\Delta_k\right)\right\>.
\end{align}
 
From now on, we divide the discussion into two cases:

\medskip

\noindent{\em Case $i)$: $R< d^\frac{3}{2}$.} In this case, it is obvious that the bound (\ref{bound}) holds. 

\medskip

\noindent{\em Case $ii)$: $R\ge d^\frac{3}{2}$.} We recall the definition of $R$ and write the condition as
\begin{align}\label{R_scaling_assumption}
R=(\mu/\sigma)^2\ge d^{\frac{3}{2}}.
\end{align}
By definition of $\Delta$ [cf.\ Eq.\ (\ref{Delta})], this implies that
\begin{align}\label{scaling-Delta}
\frac{\sigma}{\Delta}\ge d^{\frac{13}{4}}\qquad{\rm and}\qquad\frac{\mu}{\Delta}\ge d^{4}.
\end{align}
Now, let
\begin{align}\label{Kminus}
K_-=\lfloor(\mu-d^{\frac14}\cdot\sigma)/\Delta\rfloor,
\end{align}
and also define another parameter
\begin{align}\label{Kplus}
K_+:=\lceil(\mu+d^{\frac14}\cdot\sigma)/\Delta\rceil,
\end{align}
where $\lfloor\cdot\rfloor$ and $\lceil\cdot\rceil$ are the floor and ceiling functions, respectively.
Notice that $(\Delta\cdot K_\pm)/\mu\to 1$ for large enough $d$. 
Chebyshev's inequality tells us that
\begin{align}\label{bound-confidence}
\sum_{k=K_-}^{K^+}p_0^\Delta (k)\ge \int_{\mu-d^\alpha\cdot\sigma}^{\mu+d^\alpha\cdot\sigma}\d t\,f(t)\ge 1-d^{-\frac12}.
\end{align}
Therefore, for any $k<K_-$, we have
\begin{align}
P^\Delta(k)=\sum_{s=k+1}^{\infty}p_0^\Delta (s)\ge 1-d^{-\frac12}.
\end{align}
Substituting into Eq.\ (\ref{CK_and_P_tilde}), we have
\begin{align}\label{CK-bound}
\frac{1}{K_-}\le C_{K_-}^\Delta\le\frac{1}{K_-(1-d^{-\frac12})}.
\end{align}
 
Now we are ready to bound both terms on the right hand side of Eq.\ (\ref{dimension_vs_entropies_bound}). 
First, notice that 
\begin{align}
H\left(p^\Delta_k\right)&=-\sum_{l=1}^{\infty} p^\Delta_k(l)\log p^\Delta_k(l)\\
&=-\sum_{l=1}^{\infty}\frac{p^\Delta_0(k+l)}{P^\Delta(k)}\log\left(\frac{p^\Delta_0(k+l)}{P^\Delta(k)}\right).
\end{align}
Then, its average with respect to the probability distribution $\{\tilde{P}^\Delta(k)\}$ [cf.\ Eq.\ (\ref{CK_and_P_tilde})] can be bounded as
\begin{align}
\left\<H\left(p^\Delta_k\right)\right\>&=-\sum_{k=0}^{K_--1}C_{K_-}^\Delta\cdot\sum_{l=1}^{\infty} p^\Delta_0(k+l) \log\left(\frac{p^\Delta_0(k+l)}{P^\Delta(k)}\right)\\
&\le-\sum_{k=0}^{K_--1}C_{K_-}^\Delta\cdot \sum_{l=1}^{\infty}p^\Delta_0(k+l) \log p^\Delta_0(k+l)\\
&\le -\sum_{s=1}^{\infty}\left(K_-\cdot C_{K_-}^\Delta\right)p_0^\Delta(s)\log p^\Delta_0(s)\\
&\le H\left(p^\Delta_0\right)\left(1-d^{-\frac12}\right)^{-1}.
\end{align}
For $d\ge 4$ we have $1/(1-d^{-\frac12})\le 1+2d^{-\frac12}$. Under this condition, the average entropy can be bounded as
\begin{align}\label{bound_averageentropy}
\left\<H\left(p^\Delta_k\right)\right\>&\le H\left(p^\Delta_0\right)\left(1+2d^{-\frac12}\right).
\end{align}

Next, the average of $p^\Delta_k$ with respect to $\tilde{P}^\Delta(k)$ can be evaluated as
\begin{align}
\left\<p^\Delta_k\right\>(s)&=\sum_{k=0}^{K_--1}\tilde{P}^\Delta(k)\cdot p_k^\Delta(s)\nonumber\\
&=C_{K_-}^\Delta\cdot \sum_{k=0}^{K_--1}p^\Delta_0(k+s).\label{average_pk}
\end{align}
For $K_+-K_-+1\le s\le K_-$, we have, from Eqs.\ (\ref{bound-confidence}), (\ref{CK-bound}) and (\ref{average_pk}), that
\begin{align}
\left\<p^\Delta_k\right\>(s)&\ge C_{K_-}^\Delta\cdot \sum_{k=K_-}^{K_+}p^\Delta_0(k)\nonumber\\
&\ge \frac{1}{K_-}\cdot \left(1-d^{-\frac12}\right).
\end{align}


From Eq.\ (\ref{average_pk}), for every $s$ we also have $\left\<p^\Delta_k\right\>(s)\le C_{K_-}^{\Delta}$, which is smaller than $1/e$ since we assumed $d\ge 4$,\footnote{Explicitly, by Eqs. (\ref{R_scaling_assumption}), (\ref{scaling-Delta}), and (\ref{Kminus}) we have $C_{K_-}^{\Delta}\le [(\lfloor d^{4}-d^{\frac{7}{2}}\rfloor)(1-d^{-\frac12})]^{-1}$, and thus $C_{K_-}^\Delta$ is strictly smaller than $1/e$  for any $d\ge 4$.} and thus $\<p^\Delta_k\>(s)\log(1/\<p^\Delta_k\>(s))$ can be regarded as monotonically increasing.
Then, the entropy of the average distribution can be bounded as:
\begin{align}
H\left(\left\<p^\Delta_k\right\>\right)&\ge\sum_{s=K_+-K_-+1}^{K_-}\<p^\Delta_k\>(s)\log\left(\frac{1}{\<p^\Delta_k\>(s)}\right)\nonumber\\
&\ge \left(1-d^{-\frac12}\right)\cdot\left(\frac{2K_--K^+}{K_-}\right)\left(\log K_- -\log\left(1-d^{-\frac12}\right)\right)\nonumber\\
&\ge \left(1-d^{-\frac12}\right)\cdot\left(\frac{2K_--K^+}{K_-}\right)\log K_-.
\end{align}
Substituting the definitions of $K_-$ and $K_+$ [cf.\ Eqs.\ (\ref{Kminus}) and (\ref{Kplus})] into the above inequality, we get
\begin{align}
H\left(\left\<p^\Delta_k\right\>\right)&\ge \left(1-d^{-\frac12}\right)\cdot\left(1-\frac{2(d^{\frac14}\sigma+\Delta)}{\mu-d^{\frac14}\sigma-\Delta}\right)\log \left(\frac{\mu-d^{\frac14}\sigma-\Delta}{\Delta}\right).\label{bound_entropyofaveragedist0}
\end{align}
Now, we separate the leading order term with error terms (see Appendix~\ref{app-details} for more details). Eq.\ (\ref{bound_entropyofaveragedist0}) can further be bounded as
\begin{align}
H\left(\left\<p^\Delta_k\right\>\right)&\ge \left(1-d^{-\frac12}(3+2\tilde{\epsilon}_d)\right)\left(\log\left(\frac{\mu}{\Delta}\right)-d^{-\frac12}(\log e)(1+\tilde{\epsilon}'_d)\right).
\label{bound_entropyofaveragedist}
\end{align}
Here 
\begin{align}\label{error-scaling1}
\tilde{\epsilon}_d:=\frac{d^{-\frac72}+d^{-\frac12}+d^{-4}}{1-d^{-\frac12}-d^{-4}}<3d^{-\frac12}
\end{align}
since $d\ge 4$.
Combining Eq.\ (\ref{bound_averageentropy}) with Eq.\ (\ref{bound_entropyofaveragedist}) and substituting into Eq.\ (\ref{dimension_vs_entropies_bound}) yield
\begin{align}\label{bound_dimension_inter1}
\log d&\ge \left(1-d^{-\frac12}(3+2\tilde{\epsilon}_d)\right)\log\left(\frac{\mu}{\Delta}\right)-\left(1+2d^{-\frac12}\right)H\left(p^\Delta_0\right)-d^{-\frac12}(\log e)(1+\tilde{\epsilon}_d)\nonumber\\ 
&=\left(1+2d^{-\frac12}\right)\left(\log\mu-\log\Delta-H\left(p^\Delta_0\right)\right)-\tilde{\eta}_{d,R}
\end{align}
where $\tilde{\eta}_{d,R}:=d^{-\frac12}(\log e)(1+\tilde{\epsilon}_d)+d^{-\frac12}(5+2\tilde{\epsilon}_d)\log\left(\frac{\mu}{\Delta}\right)$. By Eqs.\ (\ref{Delta-mu-sigma}), (\ref{error-scaling1}) and $d\ge 4$, we have
\begin{align}
\tilde{\eta}_{d,R}&<d^{-\frac12}\left((2\log R+\log d)\left(5+6d^{-\frac12}\right)+(\log e)\left(1+3d^{-\frac12}\right)\right)\\
&<d^{-\frac12}\left(8(2\log R+\log d)+(5/2)(\log e)\right)\label{epsilon-R}.
\end{align}

Notice that $p_{0}^\Delta$ is a discrete probability distribution over $\N^*$. Denote by $\sigma^\Delta$ its standard deviation. We can show (see Appendix~\ref{app-max-entropy}) that, for $\sigma^\Delta\ge1$, the entropy of any such distribution is upper bounded as
\begin{align}\label{discrete-entropy-bound}
H\left(p^\Delta_0\right)\le \frac{1}{2}\log\left(2\pi e(\sigma^\Delta)^2\right)+\zeta_{\sigma^\Delta}
\end{align}
with  
\begin{align}\label{zeta-sigma}
\zeta_x:=\frac12(\log e)e^{-2\pi^2x^2}\left(\frac{6}{1-e^{-2\pi^2x^2}}+\frac{2\pi^2x(1+e^{-2\pi^2x^2})}{(1-e^{-2\pi^2x^2})^3}\right)
\end{align}
being an error term that vanishes exponentially fast in the $x\to\infty$ limit.
In addition, we can show (see Appendix~\ref{app-convergence}) that $\Delta\cdot\sigma^\Delta$ converges to $\sigma$ with an error
\begin{align}\label{variance-convergence} 
\left|(\Delta\cdot\sigma^\Delta)^2-\sigma^2\right|\le 4\Delta(2\mu+\Delta),
\end{align}
which implies $\sigma^\Delta\ge \sqrt{(\sigma/\Delta)^2-4(2\mu/\Delta+1)}\ge1$ as $d\ge 4$, so Eq.\ (\ref{discrete-entropy-bound}) holds. Then we have
\begin{align}
\log\Delta+H\left(p^\Delta_0\right)\le \frac{1}{2}\log\left(2\pi e(\Delta\cdot\sigma^\Delta)^2\right)+\zeta_{\sigma^\Delta}.
\end{align}
Using the convergence relation (\ref{variance-convergence}) and invoking Eq.\ (\ref{Delta-mu-sigma}), we have 
\begin{align}
\log\Delta+H\left(p^\Delta_0\right)&\le \frac{1}{2}\log\left(2\pi e \sigma^2\right)+\frac12(\log e)\left(\frac{4\Delta(2\mu+\Delta)}{\sigma^2}\right)+\zeta_{\sigma^\Delta}\\
&\le\frac{1}{2}\log\left(2\pi e \sigma^2\right)+\frac12(\log e)\left(\frac{4(2dR^2+1)}{d^2R^3}\right)+\zeta_{dR^{\frac32}/2}\\
&\le\frac{1}{2}\log\left(2\pi e \sigma^2\right)+\frac12(\log e)\left(\frac{12}{dR}\right)+\zeta_{dR^{\frac32}/2}\\
&\le\frac{1}{2}\log\left(2\pi e \sigma^2\right)+6(\log e)d^{-\frac52}+\zeta_{d^{\frac{13}{4}}/2},
\end{align}
where the second inequality comes from the monotonicity of $\zeta_x$ and $\sigma/\Delta=dR^{\frac32}\le 2\sigma^\Delta$ and the last inequality comes from the condition $R\ge d^{\frac32}$.
Then, Eq.\ (\ref{bound_dimension_inter1}) becomes
\begin{align}
\log d\ge \left(1+2d^{-\frac12}\right)\left(\log\mu-\frac{1}{2}\log(2\pi e\sigma^2)-6(\log e)d^{-\frac52}-\zeta_{d^{\frac{13}{4}}/2}\right)- \tilde{\eta}_{d,R}.
\end{align}
Finally, rearranging terms and using Eq.\ (\ref{epsilon-R}), we get 
Eq.\ (\ref{bound-explicit}).

\section{Lower bound in terms of the controllable dimension}\label{sec-intuition}

\subsection{Controllable dimension}
Theorem~\ref{theorem-main} can be applied to any signal-generating device. However, we expect that the limit can only be tight if the degrees of freedom of the device are used optimally, which may require full control of all of them (as is the case for a quantum computer). This is however not the case for many realistic devices, such as macroscopic oscillators. Indeed,  in this case it is not the overall dimension $d$ that is relevant but the effectively controlled dimension, a concept that we introduce in the following. 

First, recall that in the proof of Theorem~\ref{theorem-main}, we constructed a bipartite state $\sigma_0$ [cf.\ Eq.\ (\ref{sigma0})] between a classical control register and a quantum register. The latter consists of ``snapshots'' of the system on the way to producing a tick. The mutual information $I(C:S)_{\sigma_0}$ captures how well the generator system can be controlled by an external system. With this idea we can extend the notion of dimension to that of a controllable dimension.

For a signal generator, denote by $\{\rho_{t}\}_t$ its state orbit where each $\rho_{t}$ is an \emph{achievable} state of the generator system, i.e.\ there exists a certain time $t$ when the state of the generator matches $\rho_{t}$ before the tick production. Then, we have the following definition:
\begin{defi}[The controllable dimension]
The controllable dimension of a signal generator with state orbit $\{\rho_{t}\}_t$ is defined as
\begin{align}
d_{\rm ctrl}:=\max_{\{q_t\}}2^{I(C:S)_{\sigma(q_t)}},
\end{align}
where $\sigma(q_t)^{CS}:=\int\d t q_t|t\>\<t|^C\otimes\rho_t^S$ and $C$ is a register with an orthonormal basis labelled by $t$. 
\end{defi}

Notice that $I(C:S)$ in the above definition is a special case of the quantum mutual information (where one system is classical) \cite{holevo1973bounds}, which has widespread use as a measure of correlation, e.g., in communication theory \cite{wilde2013quantum}, thermodynamics \cite{sagawa2008second}, metrology \cite{hall2012does}, and many-body physics \cite{wolf2008area}. 
As $\sigma_0$ in Eq.\ (\ref{sigma0}) is a special case of $\sigma(q_t)$
, we have $d_{\rm ctrl}\ge 2^{I(C:S)_{\sigma_0}}$. The whole proof of bound (\ref{bound}) carries through with $d$ replaced by $d_{\rm ctrl}$, and thus we have:
\begin{theorem}\label{theo-control}
For any $c>2\pi e$ and for large enough $d_{\rm ctrl}$, any singleton generator (cf.\ Definition~\ref{defi-single-generator}) with controllable dimension $d_{\rm ctrl}$ obeys the bound 
\begin{align}
R\ge c\cdot d_{\rm ctrl}^2.
\end{align}
\end{theorem}

A perfectly controlled quantum system like a quantum computer of $n$ qubits obviously has $d_{\rm ctrl}=2^n$ and therefore the   bound suggests an exponential scaling of the sharpness in terms of $n$:
\begin{align}\label{R_quantum}
R({\rm quantum})\propto 2^{2n}.
\end{align}
Classical generators consisting of $n$ atoms, on the other hand, have infinite dimension but a controllable dimension that grows slowly with $n$.
A pendulum's state is, obviously, characterised by the position of the centre of mass. The centre of mass of $n$ atoms can be determined up to an uncertainty proportional to $1/\sqrt{n}$. Moreover, the length of the mass' trajectory is proportional to $l$, the length of the pendulum, which in turn depends on the frequency $\nu$. Ultimately, we have $d_{\rm ctrl}({\rm pendulum})\propto \sqrt{n}/\nu^2$, which suggests a sharpness that scales as
\begin{align}
R({\rm pendulum})\propto n/\nu^4.
\end{align}
The quality of the pendulum as a signal generator thus increases both with its mass (the particle number $n$) and the period of the oscillation ($1/\nu$). An analogous analysis applies to spring oscillators, whose frequencies decrease as their masses grow. 
 A similar dependence  $R({\rm planet})\propto n(M/\nu^2)^{\frac23}$ holds for planets consisting of $n$ atoms orbiting around a star of mass $M$. Therefore, both its high mass and low frequency make a planet an ideal candidate for a good signal generator.
Nevertheless, this scaling, in particular the inverse dependence on the frequency, prohibits its use for most practical applications.

Modern high-precision clocks are often based on the measurement of light rather than mechanical pendulums.
The state of light within a period is characterised by its phase, and a relative ``second'' can be defined as the time it takes the light to return to the same phase. As the phase has an uncertainty of $\sqrt{n}$ with $n$ being the photon number, we have $d_{\rm ctrl}({\rm optical\ clocks})\sim \sqrt{n}$, which is free from the curse of frequency dependence. The maximum sharpness then scales as 
\begin{align}\label{R-laser}
R({\rm optical\ clocks})\propto n.
\end{align}
Still, this corresponds to the limit obtained by measuring classical light. Modern techniques in quantum optics, e.g.\ squeezing, allow photons to be quantumly correlated. This enhances $d_{\rm ctrl}$ and consequently the sharpness quadratically, i.e.\ 
\begin{align}
R({\rm squeezed\ optical\ clocks})\propto n^2.
\end{align}
The performance, however, is still very far from a fully controlled quantum system [Eq.\ (\ref{R_quantum})].

Fixing their size (i.e.\ $n$), we conclude from the above discussion that fully controlled generators are exponentially better than  traditional generators like pendulums and oscillators, which use only collective control over its physical degrees of freedom.
We remark that the above discussion does not take into account environmental noise, which is still a dominant source of error in current devices (see also the discussion in Section~\ref{subsec-realclock} on clocks).

\section{Discussion}\label{sec-discussion}

\subsection{Rate-distortion theory}

\begin{figure}  [h!]
\begin{center}
  \includegraphics[width=0.9\linewidth]{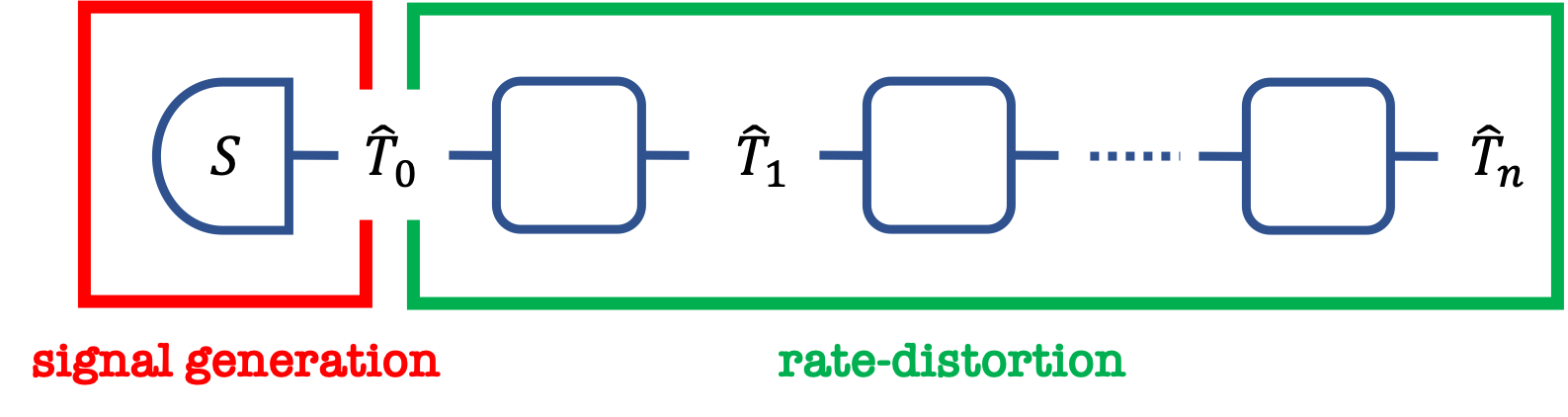}
  \end{center}
\caption{\label{fig-rate-distortion}
  {\bf Relation to   rate-distortion theory.} In information theory, a signal is generated by a source $S$ and then processed by a series of channels. Our work deals with the first step (in the red frame), i.e.\ generation, whereas   rate-distortion theory takes care of the remaining steps (in the green frame).}
\end{figure}

The rate-distortion theory founded by Shannon \cite{Shannon1948mathematical} is a  cornerstone of information theory and, in particular, the processing, compression, and storage of continuous data (see \cite[Chapter 10]{cover2012elements} for an introduction). Given an input signal, the theory deals with the tradeoff between the capacity of the channel (i.e.\ the ``rate'') used to transfer the input and the error (i.e.\ the ``distortion'') introduced by the transmission. 
However, in rate-distortion theory, the existence of an initial signal is an a priori assumption. In this sense, the bounds derived here, which are concerned with the approximate generation of a desired signal, are somewhat complementary. In a scenario that includes signal generation  and transmission, both the bound on $R$ imposed by Theorems~\ref{theorem-main} and \ref{theo-control}, as well as the rate-distortion bounds, impose fundamental limits on the overall performance of the scheme (see Figure~\ref{fig-rate-distortion}).

\subsection{Autonomous clocks}\label{subsec-clock}
Autonomous clocks \cite{rankovic2015quantum,erker2017autonomous,stupar2018performance,woods2018autonomous,woods2018quantum,yang2019accuracy} are devices that periodically output ticks to indicate the elapse of time. Their key character is that they run autonomously, without relying on any external help, e.g., for triggering measurements or calibration.

The task of single-quantum signal generation considered here is closely  
related to what an autonomous clock does. If a singleton generator's state is \emph{reset} to the initial state (instead of set to a silent state; see the discussion below Definition~\ref{defi-single-generator}) after the production of a tick, the generator can produce a sequence of ticks with i.i.d.\ intervals. The $n$-th tick then has sharpness  $R_n=n\cdot R_1$, where $R_1$ is the first-tick sharpness. 

This procedure for producing i.i.d.\ ticks corresponds to what \emph{reset clocks}, as considered in Ref.\ \cite{woods2018quantum}, do. Conversely, taking any reset clock, one can convert it into a singleton signal generator by setting the clock state to a silent state after production of the first tick. Therefore, the bound derived here applies also to reset (autonomous) clocks, and vice versa.
Furthermore, the measure $R$ considered here is identical to the measure of accuracy for a clock in \cite{woods2018quantum}. There, a particular construction of clocks called \emph{Quasi-Ideal Clocks}~\cite{woods2018autonomous} was considered and shown to have an accuracy that scales almost quadratically in its dimension $d$.
\begin{lemma}\label{lemma-quasi-ideal}[Ref.\ \cite[Theorem 2]{woods2018quantum}]
For any $x>0$, there exists $d_0(x)$ such that there is a Quasi-Ideal Clock with (first-tick) sharpness $R\ge d^{2(1-x)}$ for any $d\ge d_0(x)$. 
\end{lemma}

We now define a function $x(d)$ of $d$ as the minimal $x$ such that there exists a Quasi-Ideal Clock with $R\ge d^{2(1-x)}$. Then, we get a sequence of Quasi-Ideal Clocks with accuracy $R_{\rm Quasi-Ideal}(d)\ge d^{2(1-x(d))}$. By Lemma~\ref{lemma-quasi-ideal}, $\lim_{d\to\infty}x(d)=0$. Combining this with Theorem~\ref{theorem-main} leads to the following corollary.
\begin{cor}
There exists a sequence of Quasi-Ideal Clocks with sharpness $R_{{\rm Quasi-Ideal}}(d)$ satisfying:
\begin{align}
\lim_{d\to\infty}\frac{\log R_{\rm opt}(d)-\log R_{\rm Quasi-Ideal}(d)}{\log R_{\rm opt}(d)}=0,
\end{align} 
where $R_{\rm opt}(d)$ is the maximum first-tick sharpness of all $d$-dimensional reset clocks.  
\end{cor}
This corollary shows that the bound of Theorem~\ref{theorem-main} is asymptotically tight for large $d$.
At the same time, it asserts that Quasi-Ideal Clocks are asymptotically optimal, which addresses the open question of Ref.\ \cite{woods2018quantum}.


Another interesting observation from autonomous clocks is that quantum coherence can boost the performance. A recent result \cite[Theorem 5]{woods2018quantum} shows that the first-tick sharpness $R$ of a reset clock is upper bounded by $d$ in the absence of coherence, i.e., if the state of the clock remains diagonal during the entire process of tick production.
Translated to the general scenario of signal generation, this means that the sharpness of any incoherent generator is bounded by  
\begin{align}
R_{\rm incoh}\le d.
\end{align}
Therefore, quantum coherence enhances the performance of time signal generation quadratically.

\subsection{Implications for realistic clocks}\label{subsec-realclock}
Today even the most advanced clocks still use collectively controlled signal generators (e.g.\ a laser).
In these clocks, the dominant source of errors are imperfections and fluctuations in the Hamiltonians that govern the time evolution.
Since clocks aim to output a long sequence of equally-spaced ticks as opposed to a single quantum, changes in the Hamiltonian over time introduce an additional error in the production time $\mu$ of each tick, which is comparable to the statistical uncertainty  $\sigma$ considered here. 
In the presence of such an additional error due to imperfections, controlling more degrees of freedom does not necessarily lead to an advantage.

Nevertheless, over the past half century the uncertainty of our best clocks has been dropping from $10^{-10}$ to $10^{-18}$, mainly thanks to the development of Hamiltonian stabilising technologies \cite{derevianko2011colloquium,hinkley2013atomic,ludlow2018optical}.
It can be  expected that we will be able to find a subsystem of a large system, whose Hamiltonian is almost absolutely stable:
This is in analogy to fault-tolerance in quantum computation, where the deviation from the desired optimal behaviour can be made exponentially small (in the relevant resources, such as the number of physical qubits).

Once we have a subsystem whose Hamiltonian is fully stabilised, the bounds on accuracy of clocks described in the previous subsection become achievable. We can then build a fully quantum signal generator upon this subsystem and realise an ultra accurate clock.
For instance, to achieve a sharpness of $R=10^{20}$ (which is larger than today's record), a laser would require the same scale of photon number [cf.\ Eq.\ (\ref{R-laser})], whereas a fully controlled quantum system requires  
\begin{align}
\log\sqrt{\frac{R}{2\pi e}}\approx 32{\rm\ qubits.} 
\end{align}
Considering the speed at which quantum computers are being developed, this does not look unachievable in the mid-term future.

\subsection{Implications to the physics of spacetime}

In general relativity, clocks are necessary to operationally define the geometry of spacetime. Any fundamental bound on the accuracy of clocks can thus be translated into a limitation on how well spacetime can be mapped out (see, e.g., \cite{lloyd2012quantum,lock2017relativistic}). For this reason, the study of clocks as quantum systems within general relativity has become increasingly popular (see \cite{zych2011quantum,ruiz2017entanglement,khandelwal2019universal} for examples). 

So far, however, the role that was associated to quantum clocks in this context was mostly that of stopwatches: they measure the proper time along paths between two events, which are defined independently of the clocks. That is, the clocks only release time information upon an external trigger event, but otherwise evolve like a closed system.  Conversely, the bounds derived here are due to the requirement that time signals are generated autonomously, i.e., the clocks ``create'' events in time. In particular, Theorem~\ref{theorem-main} implies that the accuracy to which the geometry of spacetime can be determined is bounded by the dimension of the Hilbert space that describes the matter content of spacetime. Exploring this further is an interesting direction of future research.

\subsection{Quantum control}
If one wants to realise a time-dependent evolution or measure the state of a system at a given time, a signal generator is needed to control the implementation. 
Our result in particular imposes a constraint on the implementation of quantum gates. 
While this control is usually classical, it may also be interesting to apply our bounds to scenarios that involve quantum control, corresponding to a signal generator that interacts with the system coherently \cite{malabarba2015clock,woods2018autonomous}.

\subsection{Networks of signal generators}
\begin{figure}  [h!]
\begin{center}
  \includegraphics[width=0.7\linewidth]{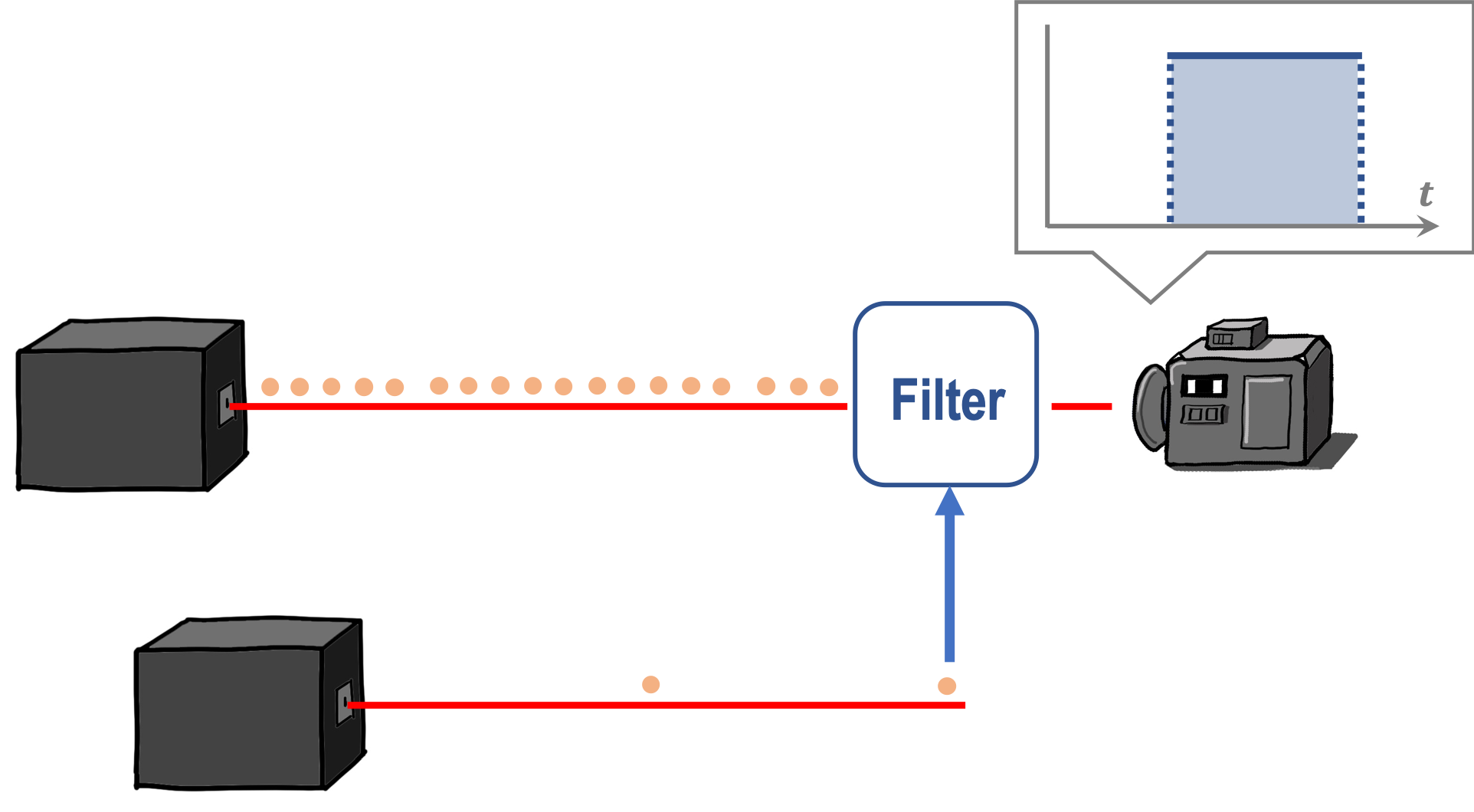}
  \end{center}
\caption{\label{fig-network}
  {\bf Continuous time signals from composite signal generators.}   The figure shows how a periodic square wave can be generated using the composition of two reset-state singleton generators (see Subsection \ref{subsec-clock}). The first generator has a much higher frequency than the second generator. The output quanta of the first generator pass through a filter, which is controlled by the second generator: On receiving a quantum from the second generator, the filter switches between on and off statuses. Finally, the output of the filter hits a detector, which displays a (continuous) square wave signal. }
\end{figure}

Signal generators can be composed and linked to a network so that they can interact with each other. 
This enables the generation of more complex time signals from basic generators (see Figure~\ref{fig-network} for an example).

By considering the whole network as a blackbox emitting one signal, we know from our bound that the sharpness of the output signal is bounded by the square of the overall dimension of all systems in the network. This imposes a limit on how much a signal can be \emph{enhanced}. In Ref.\ \cite{yang2019accuracy}, signal enhancing protocols were proposed, where the output of a signal generator can be made sharper by passing it through another generator. The performance approaches the limit predicted by our bound under a similar sharpness measure. This indicates that these protocols may be optimal.

\subsection{Modelling and simulating stochastic processes}
Stochastic models are used to make predictions based on observed data. The arguably most efficient one is the $\epsilon$-machine (see Refs.\ \cite{crutchfield1994calculi,crutchfield1989inferring,shalizi2001computational}, as well as Refs.\ \cite{gu2012quantum,mahoney2016occam,binder2018practical,ghafari2019dimensional} for  recent quantum extensions).
The most important feature of the machine is to serve as an ``Occam's razor" that compresses the observed data and cuts details irrelevant to predicting the future. The dimension of the machine is thus a key quantity. 

Complexity measures of $\epsilon$-machines  capture the cost of generating chaotic signals \cite{crutchfield1989inferring}, 
whereas for regular and periodic signals these measures tend to be very low.
The reason is that they do not take into account the cost of updating the machine state. In other words, $\epsilon$-machines are not autonomous and thus require an operator (or a clock) that updates their states.
Our result covers the cost of maintaining such an operator, thus offering a complementary perspective. 
An interesting direction for future research would be to combine these two aspects and study, for instance, autonomous $\epsilon$-machines.




\section{Conclusion}\label{sec-conclusion}
We studied the constraints that quantum physics imposes on signal generation.
We showed that the quality of a time signal crucially depends on the effectively controllable dimension of the generator. 
The approach taken here can serve as a starting point for various further investigations, e.g., on the ultimate accuracy of clocks, as discussed in Section~\ref{sec-discussion}. Understanding these fundamental limits, in turn, is crucial in any program towards a theory of spacetime in which time is defined operationally.


\begin{acknowledgements}
This work is supported by the Swiss National Science Foundation via the National Center for Competence in Research ``QSIT" as well as via project No.\ 200020\_165843. We thank Joseph Renes and Ralph Silva for discussions, and Xinhui Yang for drawing Figure~1.
\end{acknowledgements}

\bibliography{ref}
\bibliographystyle{unsrt} 

\appendix 

\section{Bounding error terms in Eq.\ (\ref{bound_entropyofaveragedist0})} \label{app-details}
Using the relations $\mu/\sigma\ge d^{\frac34}$, $\sigma/\Delta\ge d^{\frac{13}4}$ and $\mu/\Delta\ge d^{4}$ between $\mu$, $\sigma$, and $\Delta$ [see Eqs.\ (\ref{R_scaling_assumption}) and (\ref{scaling-Delta}) of the main text],
the error terms in Eq.\ (\ref{bound_entropyofaveragedist0}) can be bounded. Explicitly, first we have
\begin{align}
\frac{2(d^{\frac14}\sigma+\Delta)}{\mu-d^{\frac14}\sigma-\Delta}&=\left(\frac{2d^{\frac14}\sigma}{\mu}\right)\cdot E_{d,\Delta,\mu,\sigma}\qquad  E_{d,\Delta,\mu,\sigma}:=\left(\frac{1+d^{-\frac14}\cdot\Delta/\sigma}{1-d^{\frac14}\cdot\sigma/\mu-\Delta/\mu}\right).
\end{align}
We can bound the intermediate term $E_{d,\Delta,\mu,\sigma}$ as
\begin{align}\label{E-d-delta}
E_{d,\Delta,\mu,\sigma}\le \frac{1+d^{-\frac72}}{1-d^{-\frac12}-d^{-4}}=1+\tilde{\epsilon}_d.
\end{align}
Therefore, we have
\begin{align}
\frac{2(d^{\frac14}\sigma+\Delta)}{\mu-d^{\frac14}\sigma-\Delta}\le 2d^{-\frac12}\cdot\left(1+\tilde{\epsilon}_d\right).
\end{align}
The other term can be bounded similarly as
\begin{align}
\log \left(\frac{\mu}{\mu-d^{\frac14}\sigma-\Delta}\right)&=(\log e)\cdot\ln\left(\frac{\mu}{\mu-d^{\frac14}\sigma-\Delta}\right)\\
&\le(\log e)\left(\frac{d^{\frac14}\sigma+\Delta}{\mu-d^{\frac14}\sigma-\Delta}\right)\\
&=(\log e)\left(\frac{d^{\frac14}\sigma}{\mu}\right)E_{d,\Delta,\mu,\sigma}\\
&=\left(d^{-\frac12}\log e\right)\left(1+\tilde{\epsilon}_d\right).
\end{align}
Here $\tilde{\epsilon}_d:=(1+d^{-\frac72})/(1-d^{-\frac12}-d^{-4})-1$.

\section{Maximum entropy for discrete probability distributions [Proof of Eq.\ (\ref{discrete-entropy-bound})]}\label{app-max-entropy}
Here we prove Eq.\ (\ref{discrete-entropy-bound}) of the main text. We consider, instead, the same problem for distributions over $\Z$. Since $\N^*\subset\Z$, a bound for distributions over $\Z$ holds also for distributions over $\N^*$. 

Notice that we need a bound that works for large $\sigma$, so we simply assume $\sigma\ge 1$ (the choice is flexible), and we have  the following lemma: 
\begin{lemma}
For an arbitrary (discrete) probability distribution $q(n)$ over $\Z$, we have
\begin{align}
H\left(q(n)\right)\le\frac{1}{2}\log(2\pi e\sigma^2)+\zeta_\sigma.
\end{align}
where $\sigma^2$ is the variance of $q(n)$, assumed to be lower bounded by one, and $\zeta_\sigma$ is the exponentially vanishing error term defined in Eq.\ (\ref{zeta-sigma}). 
\end{lemma}

\noindent{\bf Proof.}
By standard Lagrangian multiplier approach, we can see that the maximum entropy distribution has the form:
\begin{align}
q^*(n)=Z_\lambda \cdot e^{-\lambda(n-\mu)^2},
\end{align}
where $\mu$ is the mean, $\lambda>0$ is a parameter and $Z_\lambda$ is a normalisation constant.
The maximum entropy can be expressed as
\begin{align}\label{appendix-max-entropy}
H(q^*)=\sum_{n}\left(\log e\cdot\lambda(n-\mu)^2-\log Z_\lambda\right)Z_\lambda e^{-\lambda(n-\mu)^2}=(\log e)\cdot\lambda\sigma^2-\log Z_\lambda.
\end{align}

The coefficients $Z_\lambda$ and $\lambda$ can be determined by the constraints:
\begin{align}\label{appendix-Zlambda-sigma-constraints1}
&\sum_n Z_\lambda \cdot e^{-\lambda(n-\mu)^2}=1\\
&\sum_n Z_\lambda(n-\mu)^2 \cdot e^{-\lambda(n-\mu)^2}=\sigma^2.\label{appendix-Zlambda-sigma-constraints2}
\end{align}

Define $g_1$ and $g_2$ as the functions 
\begin{align}
g_1(x):=e^{-\lambda(x-\mu)^2}\qquad g_2(x):=(x-\mu)^2e^{-\lambda(x-\mu)^2}
\end{align}
and $\hat{g}_1$ and $\hat{g}_2$ as their respective Fourier transforms. Explicitly, we have
\begin{align}
\hat{g}_1(x)=\sqrt{\frac{\pi}{\lambda}}e^{-\frac{\pi^2 x^2}{\lambda}+2\pi ix\mu}\qquad \hat{g}_2(x):=\frac{\sqrt{\pi}(\lambda-2\pi^2x^2)}{2\lambda^{5/2}}e^{-\frac{\pi^2x^2}{\lambda}+2\pi ix\mu}.
\end{align}
The Poisson summation formula tells us that
\begin{align}
\sum_{n\in\Z}g_x(n)=\sum_{m\in\Z}\hat{g}_x(m)
\end{align}
for both $x=1,2$. We can get a good approximation by working out the summations for the Fourier transformed functions. 
The first summation can be lower bounded as
\begin{align}
\sum_{m\in\Z}\hat{g}_1(m)\ge \hat{g}_1(0)=\sqrt{\frac{\pi}{\lambda}}
\end{align}
and also upper bounded as
\begin{align}
\sum_{m\in\Z}\hat{g}_1(m)&\le \sqrt{\frac{\pi}{\lambda}}\left(1+2\sum_{m>0}e^{-\frac{\pi^2 m^2}{\lambda}}\right)\\
&\le\sqrt{\frac{\pi}{\lambda}}\left(1+\frac{2e^{-\frac{\pi^2}{\lambda}}}{1-e^{-\frac{\pi^2}{\lambda}}}\right).
\end{align}
Similar calculations work for $\sum_{m\in\Z}\hat{g}_2(m)$. Then, we have
\begin{align}
\sum_{n\in\Z}g_1(n)&\in\left[\sqrt{\frac{\pi}{\lambda}},\sqrt{\frac{\pi}{\lambda}}\left(1+\frac{2e^{-\frac{\pi^2}{\lambda}}}{1-e^{-\frac{\pi^2}{\lambda}}}\right)\right]\\
\sum_{n\in\Z}g_2(n)&\in\left[\sqrt{\frac{\pi}{4\lambda^3}}\left(1-e^{-\frac{\pi^2}{\lambda}}\frac{\sqrt{2}\pi^2(1+e^{-\frac{\pi^2}{\lambda}})}{\sqrt{\lambda}(1-e^{-\frac{\pi^2}{\lambda}})^3}\right),\sqrt{\frac{\pi}{4\lambda^3}}\right].
\end{align}
Substituting into Eqs.\ (\ref{appendix-Zlambda-sigma-constraints1}) and (\ref{appendix-Zlambda-sigma-constraints2}), we get
\begin{align}
 \frac{1}{Z_\lambda}&\in\left[\sqrt{\frac{\pi}{\lambda}},\sqrt{\frac{\pi}{\lambda}}\left(1+\frac{2e^{-\frac{\pi^2}{\lambda}}}{1-e^{-\frac{\pi^2}{\lambda}}}\right)\right]
 \label{appendix-Zlambda-sigma-constraints3}\\
\frac{1}{\sigma^2}&\in\left[2\lambda, 2\lambda\cdot (1+\zeta'(\lambda))\right]\quad\zeta'(\lambda):=\frac{2e^{-\frac{\pi^2}{\lambda}}}{1-e^{-\frac{\pi^2}{\lambda}}}+e^{-\frac{\pi^2}{\lambda}}\frac{\sqrt{2}\pi^2(1+e^{-\frac{\pi^2}{\lambda}})}{\sqrt{\lambda}(1-e^{-\frac{\pi^2}{\lambda}})^3}.\label{appendix-Zlambda-sigma-constraints4}
\end{align}
Substituting into Eq.\ (\ref{appendix-max-entropy}) we get 
\begin{align}
H(q^*)\le\frac12\log \left(\frac{e\pi}{\lambda}\right)+\log\left(1+\frac{2e^{-\frac{\pi^2}{\lambda}}}{1-e^{-\frac{\pi^2}{\lambda}}}\right).\label{appendix-max-entropy-inter}
\end{align}
To proceed, notice that, by the assumption $\sigma\ge 1$, Eq.\ (\ref{appendix-Zlambda-sigma-constraints4}) implies that $\lambda\le 1/(2\sigma^2)\le 1/2$. Then $\zeta'(\lambda)$ in Eq.\ (\ref{appendix-Zlambda-sigma-constraints4}) is monotonically increasing, and we have $\zeta'(\lambda)\le \zeta'(1/(2\sigma^2))$ for any $\lambda$ satisfying constraint (\ref{appendix-Zlambda-sigma-constraints3}). Therefore, from Eq.\ (\ref{appendix-Zlambda-sigma-constraints4}) we get
\begin{align}
\lambda\ge\left(\frac{1}{2\sigma^2}\right)\left(1+\zeta'\left(\frac{1}{2\sigma^2}\right)\right)^{-1}.
\end{align}
Substituting back into Eq.\ (\ref{appendix-max-entropy-inter}), we get 
\begin{align}
H(q^*)&\le\frac12\log(2\pi e\sigma^2)+\frac12\log\left(1+\zeta'\left(\frac{1}{2\sigma^2}\right)\right)+\log\left(1+\frac{2}{e^{2\pi^2\sigma^2}-1}\right)\\
&\le\frac12\log(2\pi e\sigma^2)+\frac12(\log e)\left(\zeta'\left(\frac{1}{2\sigma^2}\right)+\frac{4}{e^{2\pi^2\sigma^2}-1}\right),
\end{align}
which is the desired bound.

 \qed

\section{Proof of Eq.\ (\ref{variance-convergence})}\label{app-convergence}
Here we prove Eq.\ (\ref{variance-convergence}) of the main text.
Consider an arbitrary (continuous) probability distribution $f(t)$ over $[0,\infty)$ with mean $\mu$ and variance $\sigma^2$. Denote by $p^\Delta(n):=\int_{(n-1)\Delta}^{n\Delta}f(x)\d x$ the discretisation of $f(t)$ with step length $\Delta$. Denote also by $\sigma^\Delta$ ($\mu^\Delta$) the standard deviation (the mean) of $p^\Delta$. We have the following lemma:
\begin{lemma}
In the limit of $\Delta\to0$, $\Delta\cdot\mu^\Delta$ and $\Delta\cdot\sigma^\Delta$ converge to $\mu$ and $\sigma$, respectively. The speed of convergence is determined by the following bounds:
\begin{align}
\left|\mu-\Delta\cdot\mu^\Delta\right|\le \Delta
\end{align}
and
\begin{align}
\left|\left(\Delta\cdot\sigma^\Delta\right)^2-\sigma^2\right|\le4\Delta(\Delta+2\mu).
\end{align}

\end{lemma}

\noindent{\bf Proof.}
First, we bound the gap between $\Delta\cdot\mu^\Delta$ and $\mu$ as:
\begin{align}
\left|\mu-\Delta\cdot\mu^\Delta\right|&=\left|\sum_{n=1}^{\infty}\int_0^{\Delta}\d x f(x+(n-1)\Delta)(\Delta-x)\right|\nonumber\\
&\le \Delta\sum_{n=1}^{\infty}\int_0^{\Delta}\d x f(x+(n-1)\Delta)\nonumber\\
&=\Delta.
\end{align}
Then, for the variance 
\begin{align}
\left(\Delta\cdot\sigma^\Delta\right)^2:=\sum_{n=1}^{\infty} p^\Delta(n)\Delta^2(n-\mu^\Delta)^2,
\end{align}
we have
\begin{align}
\left|\left(\Delta\cdot\sigma^\Delta\right)^2-\sigma^2\right|&=\left|\sum_{n=1}^{\infty}\int_0^{\Delta}\d x f(x+(n-1)\Delta)\left((x+(n-1)\Delta-\mu)^2-\Delta^2(n-\mu^\Delta)^2\right)\right|\nonumber\\
&\le\sum_{n=1}^{\infty}\int_0^{\Delta}\d x f(x+(n-1)\Delta)\left|(\Delta-x)+(\Delta\mu^\Delta-\mu)\right|\cdot\left|x+(2n-1)\Delta-\mu-\Delta\mu^\Delta\right|\nonumber\\
&\le 2\Delta\sum_{n=1}^{\infty}\int_0^{\Delta}\d x f(x+(n-1)\Delta) \left|x+(2n-1)\Delta-\mu-\Delta\mu^\Delta\right|\nonumber\\
&\le 2\Delta\int_{0}^{\infty}\d y f(y) 2(\Delta+|y-\mu|)\nonumber\\
&\le 2\Delta\int_{0}^{\infty}\d y f(y) 2(\Delta+y+\mu)\nonumber\\
&= 4\Delta(\Delta+2\mu).
\end{align}
\qed

\end{document}